\documentclass[journal]{IEEEtran}

\usepackage{amsmath}
\usepackage{amssymb}
\usepackage{graphicx}
\usepackage{algorithmic}
\usepackage{cite}
\usepackage{balance}
\usepackage{float}
\usepackage[caption=false,font=normalsize]{subfig}

\begin{document}

\title{Validating Navmesh using Geometry: Voxel-Based Analysis with Prioritized Exploration}

\author{
Ramesh Raghavan,
Ojas Sharma,
Sebastien Larrue,
Alan Isaac Kunder,
Aakash Sai,
Rishi Mathur\\
Ubisoft (India \& Singapore)\\
\texttt{ramesh.r@ubisoft.com, ojas.sharma@ubisoft.com, sebastien.larrue@ubisoft.com, 
alan-isaac.kunder@ubisoft.com, sai.aakash@ubisoft.com, rishi.mathur@ubisoft.com}
}

\maketitle

\begin{abstract}

Navigation mesh (Navmesh) inconsistencies affect the player experience by directly impacting the navigation systems used by non-playable characters (NPCs) in game environments. While navmeshes are generated from world geometry using well-established algorithms, environments change throughout development as terrain is adjusted and assets are moved or replaced. Due to the large scale of these environments, frequent updates, and dependencies on collision data and navmesh generation parameters, navmeshes are not always regenerated in sync with environmental changes, resulting in mismatches between the navmesh and the actual environment.

Existing automated approaches attempt to detect navigation issues using exploration agents, reinforcement learning, and other agent-driven techniques. However, since these methods typically rely on the navigation data itself or evaluate navigation behavior indirectly, they do not explicitly verify whether the navigation representation accurately reflects the true walkable space defined by the underlying geometry.

This paper presents a framework for validating navigation meshes through an independent, geometry-driven analysis of navmesh correctness. The approach reconstructs walkable space directly from environment geometry using a voxel-based representation, followed by constraint-aware traversal and connectivity evaluation. Rather than treating validation as uniform exploration, it is formulated as a prioritized search problem over the voxel space, where reinforcement learning is used to guide sampling toward regions more likely to exhibit inconsistencies. At each sampled location, reachability derived from the voxel representation is compared against reachability obtained from the navmesh via engine-level queries.

Compared to uniform and random sampling strategies, the proposed method improves validation efficiency by reducing the number of samples needed to adequately explore the validation space. Experiments across multiple large-scale open-world game environments show that the approach consistently lowers exploration effort while maintaining similar defect detection coverage. The framework runs offline within the game engine and can be integrated into automated quality assurance pipelines, allowing it to scale with large and continuously evolving environments. Since the method relies primarily on geometry and engine queries, it can be adapted across different game engines with minimal changes, making it suitable for production-level deployment.
\end{abstract}


\section{Introduction}

\IEEEPARstart{N}{avigation} meshes (navmeshes) are widely used in video games to represent traversable regions and support pathfinding for non-playable characters (NPCs). They specify which parts of an environment are walkable and how those regions connect, forming the basis for efficient navigation in complex game worlds.

In large-scale game development, environments rarely remain static. Designers and artists continuously update terrain, assets, and collision geometry as the world evolves, and navmeshes are typically regenerated through build or integration pipelines to reflect these changes.

In practice, however, keeping the navmesh perfectly aligned with the underlying geometry is not always straightforward. The scale of modern environments, frequent updates, and dependencies on collision data and generation parameters can introduce mismatches. Even small changes or slightly misaligned collision data may result in missing or incorrect walkable regions, which in turn can cause unexpected traversal paths or navigation failures for NPCs during gameplay.

\begin{figure}[t]
\centering
\includegraphics[width=\linewidth]{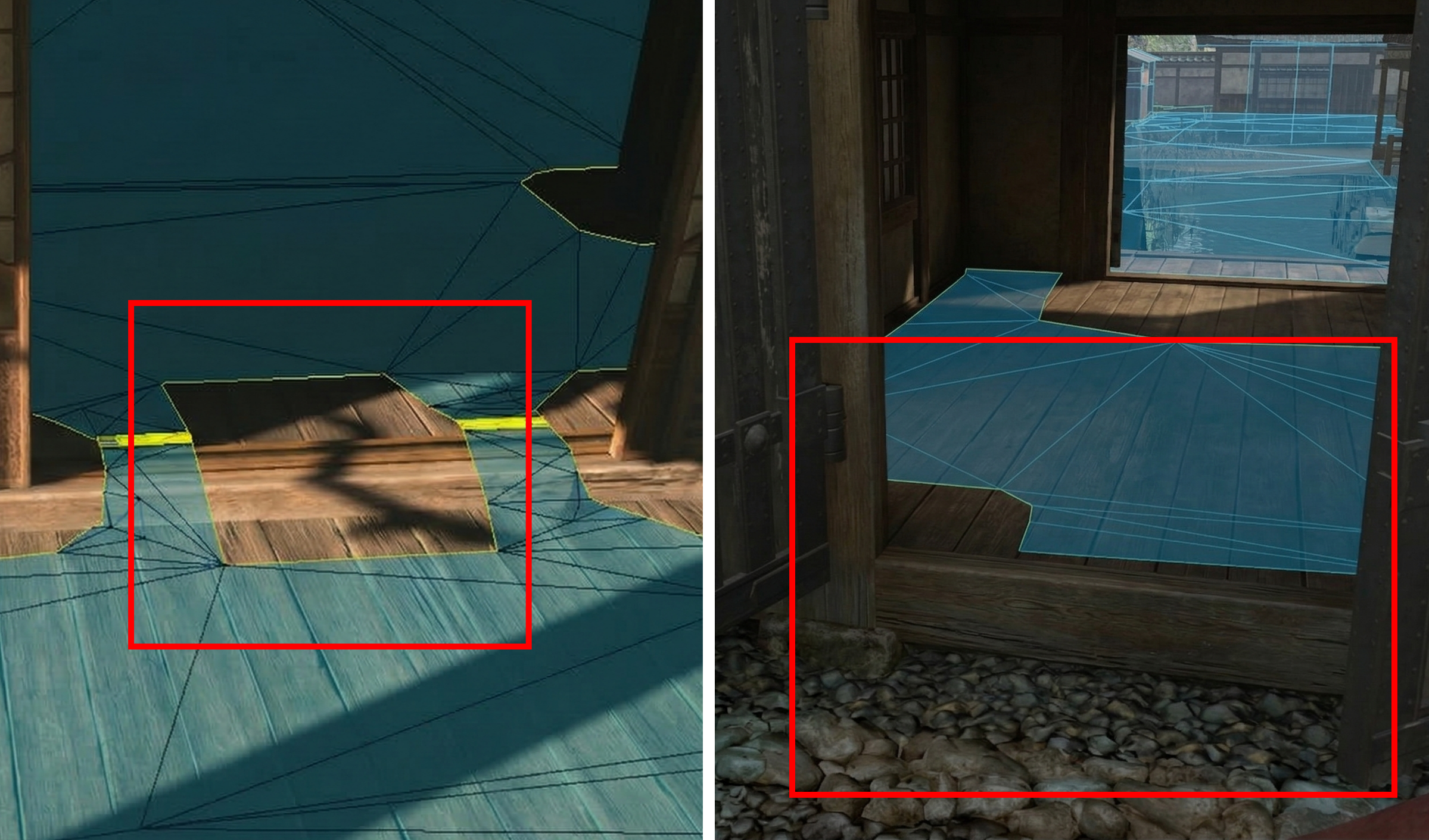}
\caption{Example of a navmesh inconsistency where the baked navmesh does not match the underlying geometry, leading to broken traversal in the highlighted region.}
\label{fig:broken_navmesh_example}
\end{figure}

Detecting these issues is often inefficient. Traditional approaches rely heavily on manual testing or agent-based exploration \cite{b1,b2,b4}, where problems are inferred indirectly from navigation behavior, such as agents getting stuck or failing to reach their targets. These methods detect issues only after they manifest at runtime, are computationally expensive, and are difficult to integrate into automated pipelines. A direct method for validating navmesh correctness independent of runtime behavior remains lacking.

To address this, navmesh validation is formulated as a geometry-consistency problem, where the objective is to verify whether the navmesh accurately represents walkable space derived from environment geometry. An offline framework is introduced that reconstructs walkable space using a voxel-based representation, followed by constraint-aware traversal and connectivity analysis.

Let $V$ denote the set of voxels derived from terrain and collision geometry, and let $W \subseteq V$ represent walkable voxels. Connectivity is represented as
\begin{equation}
G = (W, E),
\end{equation}
where $E$ defines adjacency between neighboring voxels.

Since voxelized environments can be large, exhaustive validation is expensive and uniform exploration is inefficient, as inconsistencies are often localized. Therefore, the validation is treated as a prioritized exploration problem and use reinforcement learning to guide sampling toward regions more likely to contain inconsistencies. Importantly, reinforcement learning is used for prioritization rather than simulating navigation behavior.

For each sampled location, reachability derived from the voxel model $R_{vox}(w)$ is compared with navmesh reachability $R_{nav}(w)$. A potential inconsistency is detected when
\begin{equation}
R_{nav}(w) \neq R_{vox}(w),
\end{equation}
indicating a mismatch between geometry-derived walkability and navmesh connectivity. Detected inconsistencies are refined using tolerance-based filtering and spatial clustering.

At a high level, navmesh validation is formulated as a geometry-consistency problem based on reachability agreement between two independent representations of walkable space. Instead of relying on agent behavior or assuming the correctness of the navmesh, the proposed method explicitly identifies inconsistencies as regions where geometry-derived reachability and navmesh reachability differ. This formulation enables direct verification of navigation correctness independent of runtime simulation and decouples validation from navigation execution.

The key contributions of this work are:
\begin{itemize}
\item A formulation of navmesh validation as a geometry-consistency problem, enabling direct verification of navigation correctness independent of agent-based exploration.
\item A voxel-based representation of walkable space with constraint-aware traversal, capturing walkable regions from terrain and collision data.
\item A learning-guided exploration strategy that prioritizes regions with a higher likelihood of navigation inconsistencies, improving validation efficiency over both uniform and heuristic sampling strategies.
\item An offline validation pipeline that is designed to integrate with automated QA workflows in large-scale game development.
\end{itemize}

\section{Related Work}

Research on navigation systems in games encompasses automated gameplay testing, exploration-based bug detection, navmesh construction, and coverage-driven exploration strategies. Existing studies often identify navigation issues indirectly by analyzing agent behavior during gameplay, whereas others aim to enhance navmesh generation algorithms. Despite these efforts, direct validation of navmesh correctness remains relatively unexplored. In contrast, this study directly validates navmesh correctness by comparing it to geometry-derived walkable space.

\subsection{Automated Gameplay Testing and Exploration}

Automated exploration techniques are widely used to evaluate complex game environments. A well-known example is the Go-Explore framework \cite{b1}, which enables systematic exploration of large state spaces. Similar methods have been applied to automated gameplay testing and navigation validation through exploration-driven agents \cite{b2,b7}. Reinforcement learning has also been explored to identify navigation or collision issues by observing agent behavior under different policies \cite{b4}. However, these approaches (e.g., \cite{b2}) primarily rely on behavioral signals and remain closely tied to agent policies, making direct verification of the underlying navigation representation difficult.

More broadly, automated game testing research has explored reinforcement learning, evolutionary methods, and coverage-driven exploration strategies for detecting gameplay bugs and unreachable states \cite{b21,b22}. Recent work has also investigated vision-language models for video game QA tasks such as glitch detection, visual regression testing, and bug report generation \cite{b28}. However, these approaches still rely on runtime interaction or visual inference rather than explicitly validating the correctness of the underlying navigation representation.

\subsection{Navigation Mesh Construction}

A considerable amount of research has focused on navmesh construction. Early approaches generated navmeshes through geometric decomposition or agent-based exploration \cite{b3,b5}. In modern game engines, voxelization-based pipelines are more commonly used to derive walkable regions directly from environment geometry, with systems such as Recast \cite{b14} being widely adopted.

Subsequent work has explored improvements in polygon decomposition, geometric robustness, and hierarchical navigation structures \cite{b23,b24}. Pathfinding over navmeshes has also received significant attention; for instance, Polyanya \cite{b6} introduced an optimal pathfinding algorithm specifically designed for navmesh representations. While these methods improve both navmesh generation and traversal, they generally assume that the resulting navmesh is correct, and therefore do not explicitly address its consistency with the underlying geometry.

\subsection{Coverage-Based Exploration and Path Planning}

Coverage-based exploration has been extensively studied in robotics and autonomous systems, where agents are required to systematically traverse unknown environments \cite{b8}. Classical graph traversal methods such as breadth-first search (BFS) and depth-first search (DFS) \cite{b9} are often employed to ensure complete coverage of a given space.

More advanced coverage path planning (CPP) methods have also been proposed, particularly for tasks such as inspection and mapping \cite{b8,b25}. However, in game environments, navigation relevance is rarely uniform. Certain areas tend to carry more importance due to gameplay context, such as interaction zones, patrol routes, or spawn locations. As a result, uniform coverage can be inefficient, making more targeted exploration strategies a practical necessity.

\subsection{Reinforcement Learning for Exploration}

Reinforcement learning has been widely applied to exploration and navigation in both robotics and virtual environments, with methods such as DQN and its variants showing strong performance in large state spaces \cite{b27,b15}. These approaches are commonly used for exploration, coverage optimization, and automated testing \cite{b1,b21}, where navigation issues are typically inferred indirectly through agent behavior.

However, such approaches rely on behavioral signals and are inherently dependent on agent policies, making it difficult to explicitly verify the structural correctness of the underlying navigation representation.

In this work, navmesh validation is formulated as a geometry-consistency problem, where reinforcement learning is used only to guide exploration, while validation is performed through explicit comparison between geometry-derived and navmesh-based reachability. This enables direct identification of structural discrepancies independent of agent behavior.

\section{Method Overview}
This work proposes an offline validation framework that applies geometry-driven consistency analysis within a practical game engine setting. The framework compares the engine-generated navmesh against a geometry-derived voxel-based reference of walkable space. Unlike agent-based approaches that rely on runtime behavior, the proposed method performs validation without relying on behavior-driven exploration, enabling deterministic and repeatable analysis within the game engine.

\begin{figure*}[t]
\centering
\includegraphics[width=\textwidth]{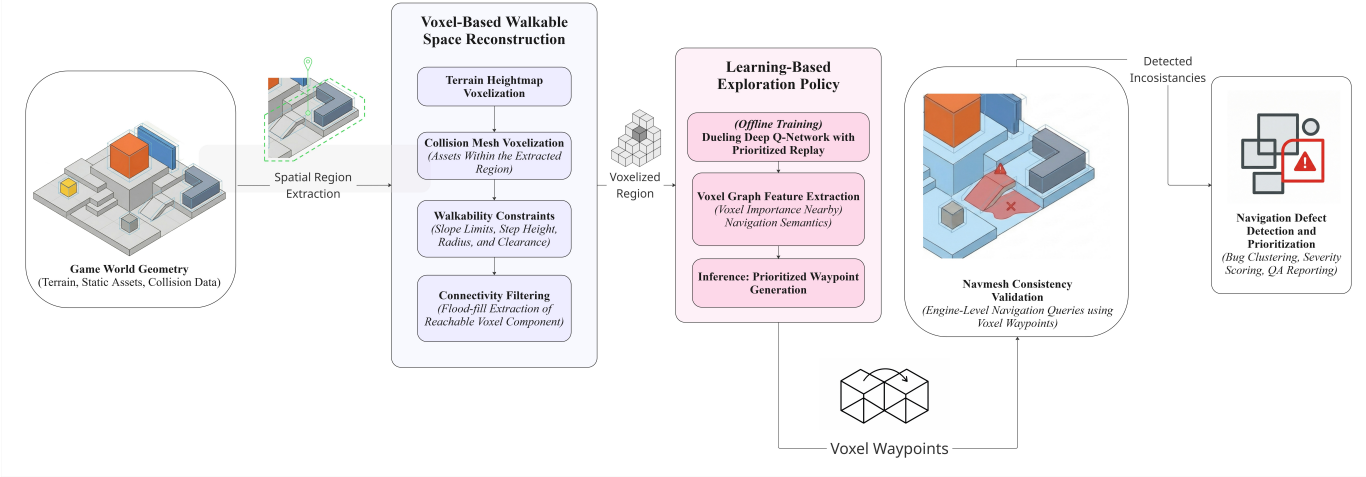}
\caption{System overview of the voxel-based navmesh validation pipeline. Walkable space is reconstructed from environment geometry, prioritized exploration selects validation locations, and engine navigation queries compare voxel-derived walkability with the generated navmesh. Detected inconsistencies are subsequently refined through tolerance-based filtering and clustering to retain only structurally significant defect regions.}
\label{fig:system_overview}
\end{figure*}

Fig.~\ref{fig:system_overview} provides an overview of how the framework operates. The proposed framework is structured around three conceptual components: (i) reconstruction of an independent geometric representation of walkable space, (ii) prioritized exploration of this space, and (iii) explicit consistency validation through reachability comparison.

\textbf{Region Extraction.}
The process begins by selecting a spatial region $R$ for validation. In practice, this is usually defined either around a seed position with a configurable radius or based on existing world partitioning schemes, such as streaming cells. Once the region is identified, all assets within it are loaded, after which non-essential elements are filtered out. Only navigation-relevant geometry is retained, including terrain surfaces and collidable objects that define traversal boundaries. This localized processing allows the method to scale to large environments without requiring full-scene evaluation. Since regions are handled independently, validation can also be parallelized across different parts of the environment, improving overall scalability.

\begin{figure}[!ht]
\centering
\subfloat[Original environment]{%
    \includegraphics[width=0.48\linewidth]{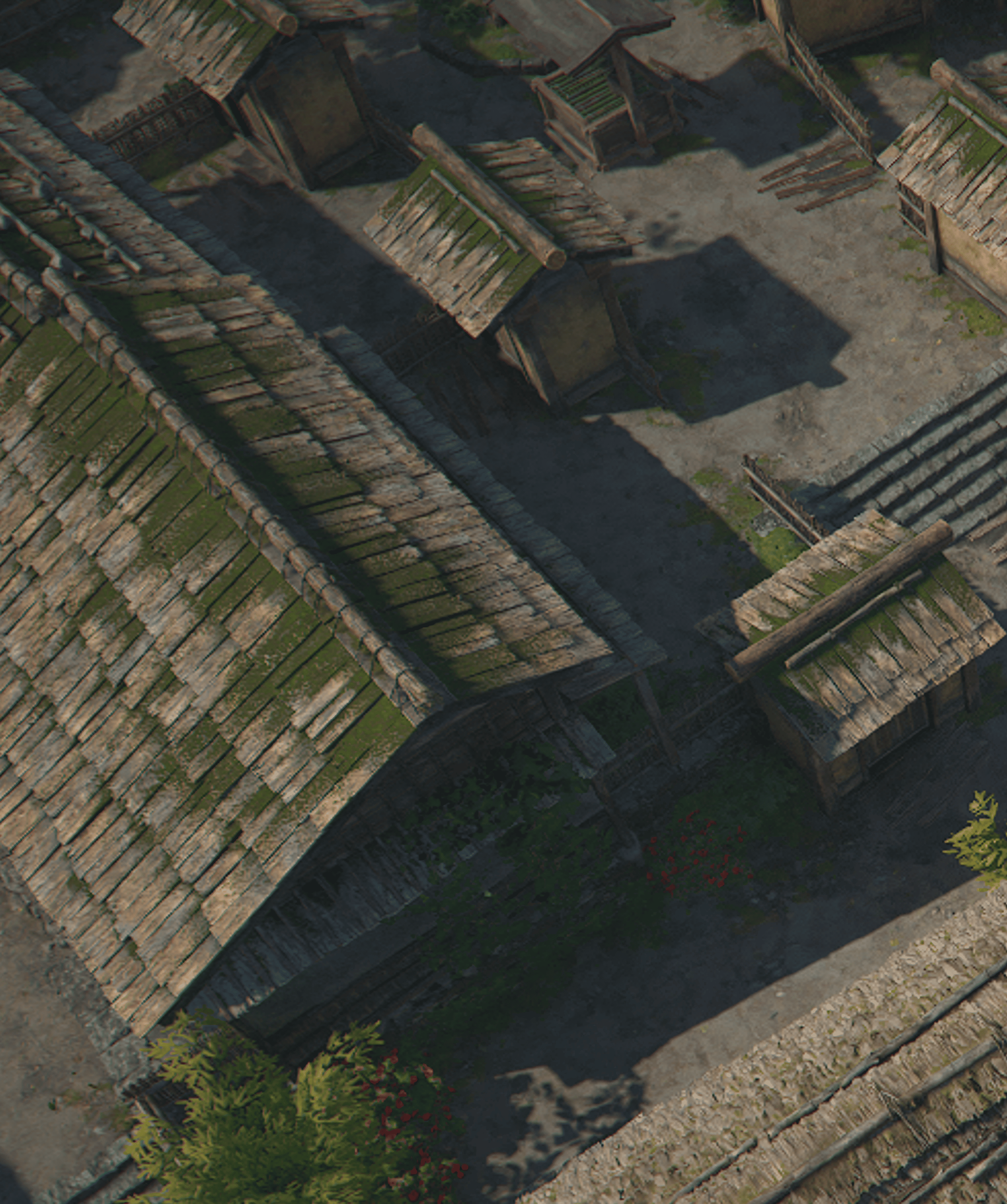}
    \label{fig:voxel_env_a}
}
\hfill
\subfloat[Voxelized walkable space]{%
    \includegraphics[width=0.48\linewidth]{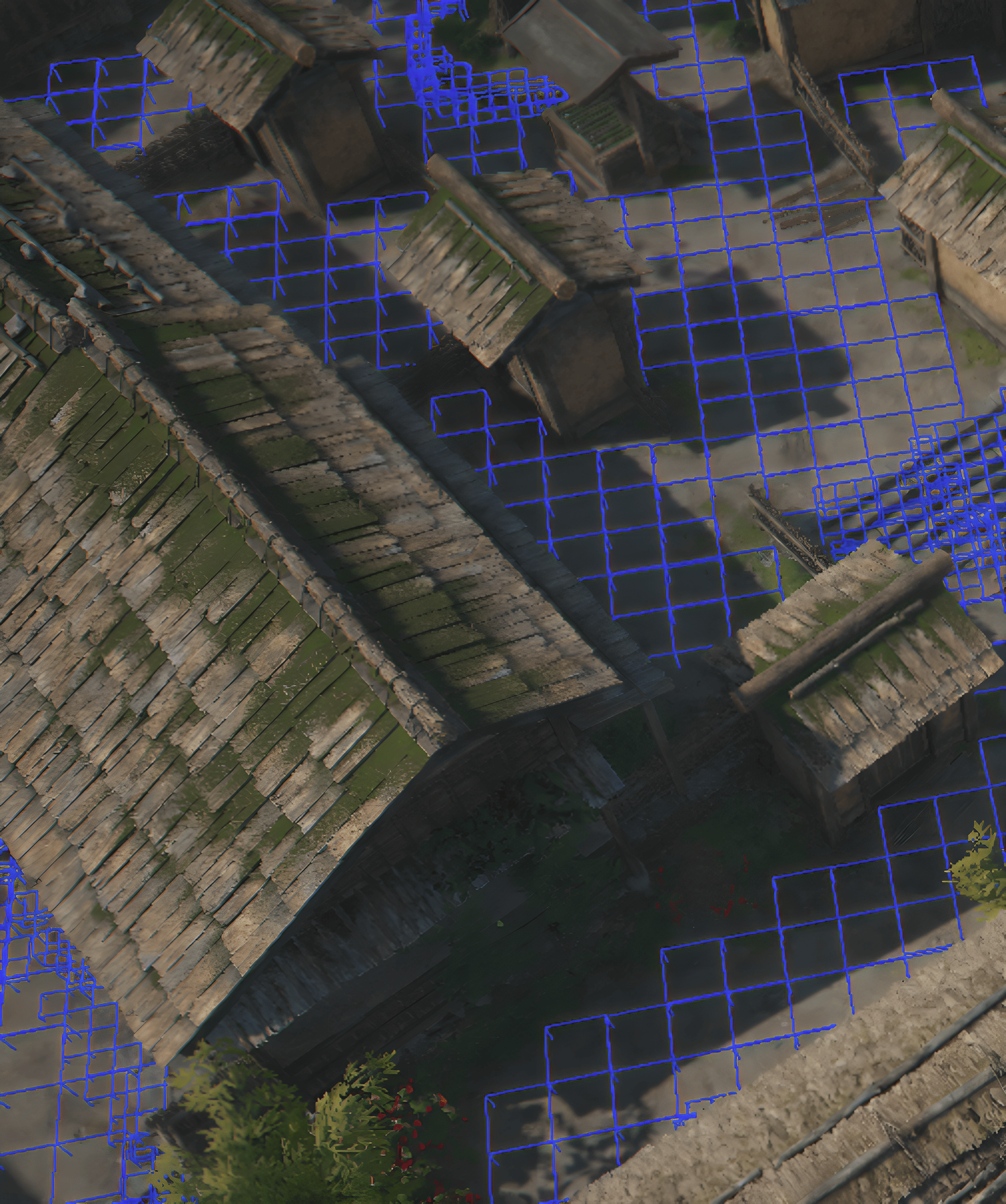}
    \label{fig:voxel_env_b}
}
\caption{Original environment and corresponding voxelized walkable space reconstruction. The voxel representation captures terrain structure, collision geometry, and navigable connectivity, serving as a geometry-driven reference for validating navmesh consistency. Non-walkable elevated structures (e.g., rooftops) are excluded, as vertical traversal is not supported.}
\label{fig:voxel_scenes}
\end{figure}

\textbf{Voxel-Based Walkable Space Reconstruction.}
Given the extracted region, an approximation of walkable space is reconstructed using a voxel-based representation that acts as an independent model derived from terrain and collision geometry. In this representation, the environment is divided into small 3D grid cells (voxels), each encoding local occupancy and traversal feasibility. An example of this reconstruction is shown in Fig.~\ref{fig:voxel_scenes}.

To build this representation, terrain is sampled through heightmaps, while assets are voxelized based on their collision meshes. Each voxel is then evaluated using a constraint-aware traversal formulation that accounts for navigation constraints such as slope limits, step height tolerance, agent radius, and clearance. Voxels that fail to satisfy these constraints are discarded, leaving only those that meet traversal requirements. A flood-fill procedure is subsequently applied to the remaining voxels to identify connected regions that are reachable from a user-defined seed location, resulting in a graph-based representation of walkable space.

This voxel-based model provides a consistent, geometry-driven approximation of walkable space that remains independent of the navmesh. The regular discretization simplifies evaluation of navigation constraints and connectivity, avoiding the need for complex polygonal processing. As a lightweight representation intended for offline validation, it does not reconstruct a full navmesh, ensuring that validation remains independent of the assumptions and processing steps used in navmesh generation. The complete reconstruction pipeline is illustrated in Section~\ref{sec:voxel_section}.

\textbf{Learning-Guided Exploration.}
For large voxelized environments, the number of candidate locations can be substantial, making exhaustive validation computationally expensive. Moreover, navigation inconsistencies are often localized rather than uniformly distributed. This makes uniform or heuristic sampling inefficient, as large portions of the space may not contain relevant inconsistencies. To address this, exploration of the voxelized walkable space is treated as a learning-guided process, where reinforcement learning is used to prioritize regions that are more likely to contain navigation inconsistencies. A Double Deep Q-Network (DDQN) formulation is used, as it is well-suited for the discrete action space over candidate regions. In addition, a prioritized experience replay buffer is employed to bias learning toward more informative samples.

The trained policy is integrated into the engine through an internal ML inference framework, allowing exploration decisions to be evaluated efficiently during validation without relying on external runtime dependencies. In this setup, reinforcement learning is used to steer sampling toward regions that exhibit higher geometric complexity or are more likely to contain inconsistencies, rather than to simulate navigation behavior directly. This targeted sampling improves overall validation efficiency.

\begin{figure}[t]
\centering
\footnotesize
\begin{tabular}{p{0.95\linewidth}}
\hline
\textbf{Algorithm 1: Voxel-Based Navmesh Validation with Post-Processing} \\
\hline
\begin{algorithmic}[1]
\REQUIRE World geometry $G$, navmesh $N$, voxel resolution $s$
\ENSURE Set of filtered navigation inconsistencies $\mathcal{I}_\epsilon$

\STATE Extract spatial region $R \subseteq G$

\STATE Generate terrain voxels $v_i=(x_i,y_i,h(x_i,y_i))$
\STATE Voxelize collision meshes where $v_i \in V_{occ}$ if $B_i \cap M \neq \emptyset$
\STATE Construct voxel set $V = V_{terrain} \cup V_{occ}$

\STATE Apply walkability constraints:
\STATE \quad slope $\theta=\arccos(n\cdot u)$,
\STATE \quad step height $|h_i-h_j|\le h_{step}$,
\STATE \quad radius clearance $d_{obs} \ge r_{agent}$,
\STATE \quad vertical clearance $H_{free}\ge H_{agent}$

\STATE Construct walkable voxel graph $G_v=(V_w,E)$ where $(v_i,v_j)\in E$ if $\|p_i-p_j\|\le r$

\STATE Train RL exploration policy $\pi_\theta$

\FOR{each exploration episode}
    \STATE Sample waypoint sequence $W=\{v_0,v_1,...,v_n\}$ using $\pi_\theta$
    \FOR{each waypoint $w \in W$}
        \STATE Compute voxel reachability $R_{\text{vox}}(w)$
        \STATE Query navmesh reachability $R_{\text{nav}}(w)$
        \IF{$(w \in R_{\text{vox}}) \neq (w \in R_{\text{nav}})$}
            \STATE $\mathcal{I} \leftarrow \mathcal{I} \cup \{w\}$
        \ENDIF
    \ENDFOR
\ENDFOR

\STATE Apply tolerance filtering to $\mathcal{I}$ to obtain $\mathcal{I}_\epsilon$
\STATE Compute connected components $\mathcal{C}$ of $\mathcal{I}_\epsilon$
\STATE Remove components $C_i$ where $|C_i| < \tau$
\STATE Update $\mathcal{I}_\epsilon$ using remaining components

\STATE \textbf{return} $\mathcal{I}_\epsilon$

\end{algorithmic} \\
\hline
\end{tabular}
\caption{Algorithm for geometry-driven navmesh validation using voxel-based walkability reconstruction, with tolerance-aware filtering and clustering for robust inconsistency detection.}
\label{alg:navmesh_validation}
\end{figure}

\textbf{Navigation Mesh Consistency Validation.}
For each sampled spatial location selected through the learning-guided exploration process, engine-level navigation queries are performed to evaluate reachability on the generated navmesh. The result is then compared with reachability derived from the voxel-based representation at the same location. If the two do not align, the location is flagged as a potential inconsistency. Detected inconsistencies are further refined through spatial post-processing, where tolerance-based filtering and connected-component clustering are applied to remove small or isolated artifacts and retain only structurally significant defect regions. This direct comparison allows identification of discrepancies between geometry-derived walkability and navmesh connectivity, enabling direct verification of navmesh consistency with the underlying geometry without relying on automated agent behavior.

The overall validation process operates offline within the game engine, eliminating the need for runtime agent simulation or NPC spawning. This makes the framework suitable for integration into automated QA workflows and continuous validation pipelines. The voxelized representation enables independent processing of spatial regions, allowing validation to be parallelized and scheduled efficiently across large and evolving environments.

\section{Voxel-Based Walkability Representation}
\label{sec:voxel_section}

To validate navmesh consistency with underlying geometry, the proposed system reconstructs an independent approximation of walkable space. Terrain and asset collision geometry are voxelized into a discrete representation, enabling localized evaluation of walkability.

Each voxel encodes its spatial position along with local traversal feasibility, bringing together surface geometry and collision constraints in a single representation. This results in a consistent and efficient approximation of navigable regions, without the need for complex geometric processing.

\begin{figure}[t]
\centering
\includegraphics[width=\linewidth]{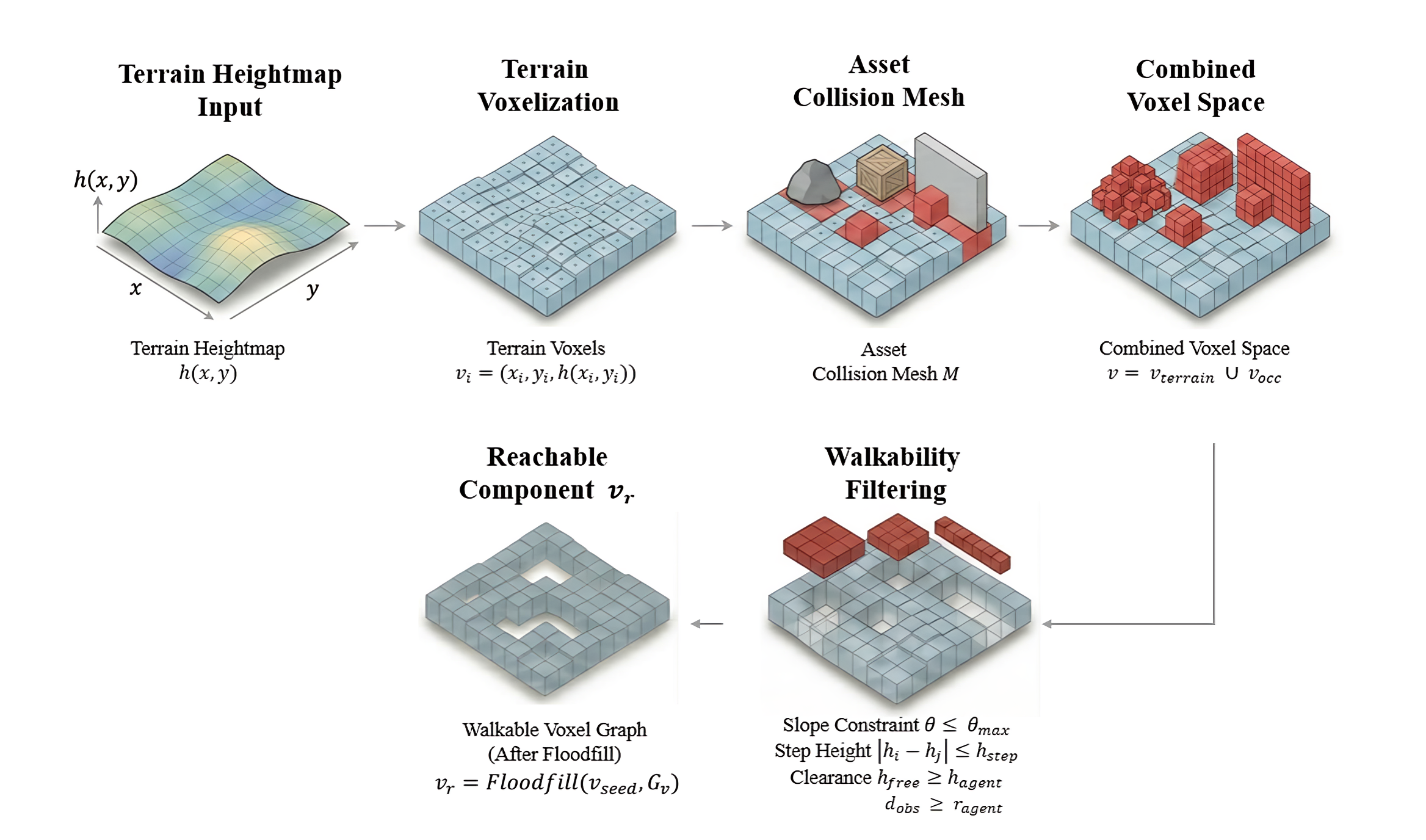}
\caption{Voxel-based reconstruction of walkable space. Terrain and collision geometry are voxelized to generate candidate voxels, after which geometric constraints and connectivity filtering remove infeasible regions. The remaining voxels approximate navigable space used for validation.}
\label{fig:voxel_connectivity}
\end{figure}

The reconstruction pipeline consists of terrain voxelization, asset collision voxelization, walkability classification, and connectivity analysis (Fig.~\ref{fig:voxel_connectivity}), forming a geometry-driven reference for identifying navigation inconsistencies.

\subsection{Terrain Voxelization}

Terrain geometry is discretized by sampling the engine heightmap, where each terrain coordinate $(x,y)$ is associated with a height value $h(x,y)$ representing the surface. For each sampled coordinate, a voxel corresponding to the terrain surface is generated as

\begin{equation}
v_i = (x_i, y_i, h(x_i,y_i)).
\end{equation}

Here $v_i$ denotes the center of the voxel representing the terrain surface at position $(x_i,y_i)$. The resulting voxel set $V_{terrain}$ represents discrete samples of the terrain surface and serves as the initial set of candidate locations. This provides a dense and structured approximation of the terrain, forming a basis for subsequent walkability evaluation.

\subsection{Asset Collision Voxelization}

Assets within the selected region are voxelized using their collision meshes, as navigation is governed by collision geometry rather than visual appearance. Assets explicitly marked to be excluded from navmesh generation (e.g., non-collidable or decorative objects that do not affect navigation) are omitted from voxelization. For the remaining assets, voxel–triangle intersection tests are performed against their collision meshes. Spatial filtering is applied to limit computations to nearby triangles, allowing efficient identification of voxels that intersect solid geometry. A voxel $v_i$ is classified as \textit{occupied} if its bounding volume intersects any triangle of the collision mesh:

\begin{equation}
v_i \in V_{occ} \quad \text{if} \quad B_i \cap M \neq \emptyset,
\end{equation}

where $B_i$ denotes the axis-aligned bounding volume of voxel $v_i$ and $M$ represents the collision mesh of the asset. The set of occupied voxels $V_{occ}$ represents discrete samples of collision geometry within the environment. The complete voxel representation used for walkability analysis is defined as:

\begin{equation}
V = V_{terrain} \cup V_{occ}.
\end{equation}

Together, $V_{terrain}$ and $V_{occ}$ provide a unified representation of traversable surfaces and collision geometry within the environment.

\subsection{Voxel Representation and Spatial Indexing}

Each voxel $v_i$ is defined by its center $p_i$ and resolution $s$, i.e., $v_i = (p_i, s)$, with spatial extent $B_i = [p_i - \tfrac{s}{2},; p_i + \tfrac{s}{2}]$.

Voxels are indexed spatially using a KD-tree \cite{b20}, enabling efficient nearest-neighbor queries, where $v_{nn} = \arg\min_{v_i \in V} |p - p_i|$.

The voxel resolution $s$ introduces a trade-off between geometric fidelity and computational cost. Smaller voxels capture finer details but increase memory usage and computation time, whereas larger voxels improve scalability at the expense of precision. In practice, $s$ is selected based on the scale of the environment and the traversal constraints of the agent.

Due to discretization, fine structures may not always be fully represented, which can introduce local errors. To address this, validation is performed at the connectivity level rather than at individual voxels, reducing sensitivity to small, isolated artifacts.

\subsection{Walkability Classification}

Each voxel is evaluated for traversal feasibility based on local geometric conditions, rather than relying on precomputed navigation data. The goal here is to approximate feasible movement regions directly from geometry, following constraints commonly used in navigation pipelines \cite{b5,b14}.

Walkability between neighboring voxels $v_i$ and $v_j$ is determined by evaluating slope, step height, agent radius clearance, and vertical clearance conditions. These parameters are taken directly from the navigation configuration of the game engine (e.g., agent radius, step height, and slope limits), ensuring consistency with in-game traversal behavior while allowing adaptation to different game-specific requirements.

Surface slope is computed using the local surface normal $n$ and the world up vector $u$:

\begin{equation}
\theta = \arccos(n \cdot u).
\end{equation}

A voxel is considered walkable if the following conditions are satisfied:

\begin{equation}
\begin{aligned}
\theta &\leq \theta_{max}, \\
|h_i - h_j| &\leq h_{step}, \\
d_{obs} &\ge r_{agent}, \\
H_{free} &\ge H_{agent}.
\end{aligned}
\end{equation}

Voxels that do not meet these conditions are classified as non-walkable and excluded from further analysis. This ensures that only geometrically feasible regions, aligned with agent traversal constraints, are retained for subsequent connectivity evaluation.

\subsection{Walkable Voxel Connectivity Graph}

The resulting walkable voxels are then connected to form a graph representation of navigable space,

\begin{equation}
G_v = (V_w, E),
\end{equation}

where $V_w$ denotes the set of walkable voxels and $E$ represents adjacency relationships between them. Two voxels are considered adjacent when their centers satisfy

\begin{equation}
(v_i, v_j) \in E \iff |p_i - p_j| \leq r,
\end{equation}

where $r$ defines the neighborhood radius determined by the voxel resolution.

This graph provides a geometry-derived connectivity model that is independent of the navmesh. Unlike navmesh connectivity, which depends on generation heuristics and simplifications, the voxel graph captures local traversability relationships directly from geometry. Discrepancies between the voxel graph and navmesh connectivity directly correspond to inconsistencies between the navigation mesh and the underlying geometry.

\subsection{Reachable Component Extraction}

The voxelization process may introduce small, isolated walkable regions that are not part of the main connected navigable space. To eliminate these artifacts, reachability filtering is applied using a flood-fill traversal over the voxel graph, retaining only the primary connected component.

Given a seed voxel $v_{seed}$, the reachable voxel set is defined as

\begin{equation}
V_r = \text{FloodFill}(v_{seed}, G_v).
\end{equation}

Only voxels in the connected component $V_r$ are retained for validation. This ensures that the analysis is restricted to regions that are reachable within the environment, excluding isolated artifacts introduced by voxelization.

\section{Learning-Guided Prioritized Exploration}

Validating navmesh correctness requires exploring the voxel graph $G_v$. In large-scale environments, this graph can include hundreds to thousands of voxels, making exhaustive traversal computationally expensive. Conventional strategies such as BFS, DFS, or uniform random sampling ensure coverage, but they distribute exploration effort uniformly, without accounting for the uneven distribution of navigation inconsistencies.

In practice, these inconsistencies are often concentrated in gameplay-relevant regions, such as interaction zones, patrol paths, and narrow passages. Because of this, uniform exploration typically requires near-complete traversal to achieve high detection rates, leading to inefficient use of computational resources.

To address this limitation, navmesh validation is formulated as a learning-guided exploration problem, where the objective is to prioritize regions that are more likely to contain inconsistencies under a constrained exploration budget.

\subsection{Why Reinforcement Learning?}

Conventional exploration strategies do not adapt to differences in environment structure or gameplay relevance. Heuristic methods introduce prioritization through predefined rules, but they remain static and often require manual tuning, which limits their ability to generalize across diverse environments.

Reinforcement learning addresses this by learning an adaptive exploration policy directly from interaction with the voxelized environment. The agent learns to prioritize regions based on spatial patterns such as geometric complexity, semantic importance, and exploration history, enabling more efficient sampling in large and non-uniform environments. This allows the policy to focus exploration on regions that are more likely to contain navigation inconsistencies, rather than distributing effort uniformly.

Importantly, reinforcement learning is used only for exploration prioritization and does not influence validation correctness, which is determined solely by geometry-based reachability comparison.

\begin{figure}[!t]
\centering
\includegraphics[width=\linewidth]{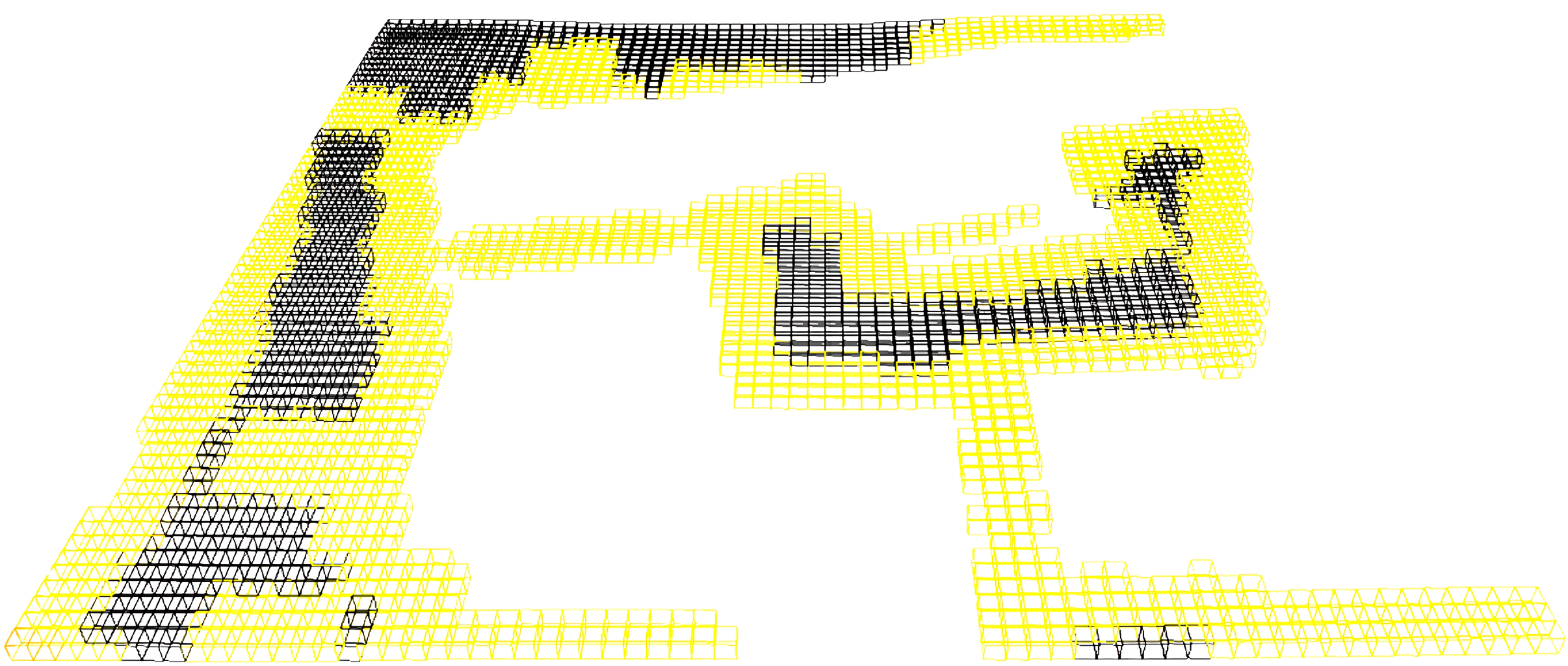}
\caption{Visualization of the voxel-based walkable space obtained through the
proposed voxelization and walkability reconstruction process in one of the
evaluated game environments. Semantically important regions are highlighted
in yellow, illustrating their uneven spatial distribution and motivating prioritized exploration over uniform sampling.}
\label{fig:voxel_importance}
\end{figure}

\subsection{Problem Formulation}

Prioritized exploration over the voxel graph $G_v = (V_w, E)$ is modeled as a finite Markov Decision Process (MDP). At each timestep $t$, the agent observes a state $s_t$ capturing local voxel context and selects an action $a_t$ corresponding to the next voxel to explore.

The objective is to learn a policy $\pi_\theta$ that maximizes the expected cumulative reward:
\begin{equation}
J(\pi_\theta) = \mathbb{E} \left[ \sum_{t=0}^{T} \gamma^t R_t \right],
\end{equation}
where $\gamma \in [0,1]$ is the discount factor. The reward formulation biases exploration toward defect-relevant regions while penalizing redundant traversal.

\subsection{State and Action Representation}

At each timestep, the agent observes a state $s_t \in \mathbb{R}^d$ encoding local geometric structure, semantic importance, and exploration status. In this formulation, the state is represented as a fixed-length feature vector capturing (i) the importance of the current voxel, (ii) exploration progress, (iii) statistics of neighboring voxels, and (iv) directional guidance toward unexplored important regions.

Specifically, the state captures normalized measures of the current voxel’s importance, accumulated coverage over important regions, and both the average and maximum importance of neighboring voxels, along with the number of unvisited important neighbors. Additional features reflect exploration dynamics, including a stagnation indicator (measured as steps since the last reward) and a directional vector pointing toward the nearest unvisited important voxel, along with its distance. Taken together, these features provide a mix of local context and broader guidance, enabling more efficient exploration.

The action space consists of discrete movement directions over neighboring voxels:
\begin{equation}
A = \{N, NE, E, SE, S, SW, W, NW\}.
\end{equation}

The next voxel is selected from the neighborhood $\mathcal{N}(v_i)$ based on directional alignment:
\begin{equation}
v_{next} =
\arg\max_{v_j \in \mathcal{N}(v_i)}
(\vec{d}a \cdot \vec{d}{v_i \rightarrow v_j}).
\end{equation}

\subsection{Semantic Importance Modeling}

Each voxel is assigned a semantic importance score:
\begin{equation}
I(v) = \sum_{k \in \mathcal{N}(v)} w_k,
\end{equation}
where $\mathcal{N}(v)$ represents the set of navigation-relevant elements within a local spatial neighborhood of voxel $v$, such as patrol paths, spawn points, or interaction zones, and $w_k$ are configurable weights that capture their relative importance. This formulation aggregates nearby gameplay signals into a continuous importance measure, which in turn biases exploration toward regions more likely to exhibit navigation inconsistencies.

\subsection{Reward Function}

The reward function is designed to encourage exploration of important regions while discouraging redundant traversal:
\begin{equation}
R_t = I(v_t) - \lambda_{step} - P_{revisit},
\end{equation}
where $I(v_t)$ provides a positive signal for visiting semantically important voxels, $\lambda_{step}$ introduces a small penalty per step to promote efficiency, and $P_{revisit}$ penalizes revisiting previously explored voxels. Together, these terms balance the need to explore high-value regions with the goal of maintaining efficient coverage.

Exploration quality is evaluated using semantic coverage:
\begin{equation}
Coverage =
\frac{\sum_{v \in V_{visited}} I(v)}
{\sum_{v \in V_w} I(v)},
\end{equation}
which measures the proportion of total semantic importance covered during exploration. This shifts the focus from uniform spatial coverage to coverage of regions that are more relevant for detecting inconsistencies.

\subsection{Policy Network}

The exploration policy is implemented using a Double Deep Q-Network (DDQN) \cite{b15}, which is well-suited to the discrete action space considered here. The DDQN formulation helps reduce overestimation bias and supports more stable learning, particularly when making discrete exploration decisions in large state spaces. The Q-function estimates expected return:
\begin{equation}
Q(s,a) = \mathbb{E}[R_t + \gamma \max_{a'} Q(s', a')],
\end{equation}
guiding movement toward voxels with higher expected validation value.

\begin{figure}[!ht]
\centering
\includegraphics[width=0.85\linewidth]
{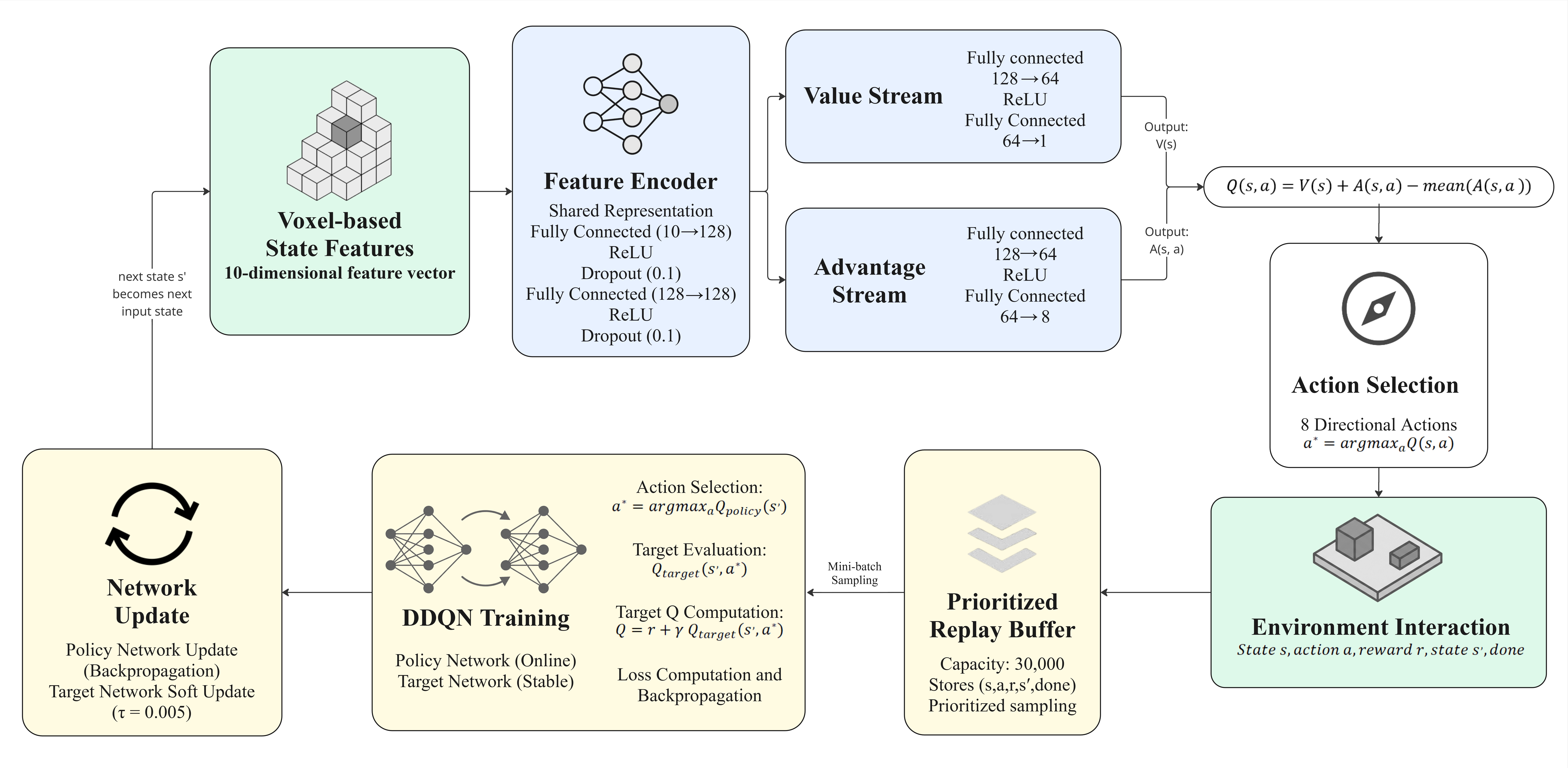}
\caption{DDQN-based exploration framework for prioritizing regions with higher likelihood of navigation inconsistencies.}
\label{fig:double_dqn}
\end{figure}

\subsection{Prioritized Experience Replay}

Prioritized experience replay \cite{b10} improves sample efficiency by biasing training toward more informative transitions. Instead of sampling experiences uniformly, transitions are prioritized based on their temporal-difference (TD) error:
\begin{equation}
\delta_t = \left| R_t + \gamma \max_{a'} Q(s_{t+1}, a') - Q(s_t, a_t) \right|.
\end{equation}

Transitions with larger TD error are sampled with higher probability, typically defined as:
\begin{equation}
P(i) = \frac{\delta_i^\alpha}{\sum_j \delta_j^\alpha},
\end{equation}
where $\alpha$ controls the degree of prioritization.

In the context of exploration over the voxel graph, this enables the agent to focus on transitions associated with semantically important or structurally complex regions, where navigation inconsistencies are more likely to occur. By emphasizing such informative experiences during training, prioritized replay accelerates convergence and improves the stability of the learned exploration policy.

\subsection{Learned Exploration Strategy}

The learned policy produces a non-uniform exploration strategy that:
\begin{itemize}
\item prioritizes regions with high semantic importance;
\item reduces redundant exploration;
\item progressively expands coverage toward high-value regions.
\end{itemize}

This results in more efficient detection of navigation inconsistencies compared to uniform traversal.

\subsection{Comparison with Baselines}

The proposed method is evaluated against uniform random sampling, exhaustive traversal strategies (BFS and DFS), and heuristic importance-based sampling. Performance is measured in terms of exploration efficiency and detection capability, with quantitative results presented in Section~\ref{sec:results}.

\section{Experimental Evaluation}
\label{sec:results}

The proposed framework is evaluated on multiple large-scale game environments to assess detection rate, exploration efficiency, and scalability.

\begin{table*}[!t]
\centering
\small
\caption{Representative test environments across different scene types and engines, illustrating variations in spatial scale, asset density, voxel counts, as well as reconstruction and validation times.}
\label{tab:env_stats}
\resizebox{\textwidth}{!}{
\begin{tabular}{lcccccc}
\hline
Environment Type & Engine & Area (m$^2$) & Voxels & Assets & Reconstruction Time (min) & Validation Time (min) \\
\hline
Structured Environments & Unity & 8k & 183k & 3k & 0.3 & 0.9 \\
Terrain-Dominant Regions & Anvil & 40k & 422k & 20k & 0.6 & 3.1 \\
Hybrid Regions & Anvil & 46k & 617k & 46k & 1.0 & 7.3 \\
\hline
\end{tabular}
}
\end{table*}

\subsection{Dataset and Setup}

Experiments are carried out on multiple production-scale environments extracted from AAA game builds (Table~\ref{tab:env_stats}). These environments span both \textbf{Unity} and Ubisoft's \textbf{Anvil Engine}, covering a range of navigation systems and scene scales. Reported results are averaged across all evaluated environments.

Voxel density varies slightly across environments, mainly due to differences in terrain complexity and asset distribution. Reconstruction time generally scales with voxel count, whereas validation time is influenced by the number of exploration samples and the cost of navmesh queries.

\subsection{Voxel Reconstruction Across Scene Types}

Voxel reconstruction typically completes within approximately 15--60 seconds per region, while full validation (including exploration and navmesh queries) completes within approximately 1--7 minutes depending on environment scale, enabling practical validation of large-scale environments within reasonable processing time.

\subsection{Evaluation Metrics}

Performance of exploration strategies is evaluated over the voxel-based representation using the following metrics:

\begin{itemize}
\item \textbf{Detection Rate (\%)}: Fraction of navigation inconsistencies detected, measured relative to the set of defects obtained via exhaustive sampling, which serves as a proxy for ground truth. Detection rate is evaluated under a bounded exploration budget, and therefore reflects practical detection performance rather than full coverage.
\item \textbf{Coverage (\%)}: Importance-weighted coverage of relevant regions
\item \textbf{Efficiency}: Number of samples required to reach a target importance-weighted coverage (set to 85\% in the experiments)
\item \textbf{False Positives (\%)}: Fraction of detected inconsistencies that do not correspond to true navigation errors after tolerance-based filtering to remove minor discrepancies arising from voxel discretization and numerical precision, followed by spatial clustering.
\end{itemize}

All methods are evaluated under a fixed exploration budget to ensure fair comparison, making efficiency (samples required) the primary differentiating factor.

\subsection{Validation and Exploration Evaluation}

To evaluate the effectiveness of the proposed geometry-driven validation framework, validation performance is analyzed over the voxelized walkable space under different exploration strategies, including random sampling, exhaustive traversal (BFS/DFS), heuristic importance-based sampling, and learning-guided exploration.

\begin{table}[!ht]
\centering
\small
\caption{Performance comparison across exploration strategies}
\label{tab:main_results}
\resizebox{\columnwidth}{!}{
\begin{tabular}{lcccc}
\hline
Method & Det. Rate & Cov. & Samples (\%) & FP \\
\hline
Random & 87.9 & 80.8 & 100\% & 9.8 \\
DFS & 90.9 & 85.0 & 100\% & 7.2 \\
BFS & 91.3 & 85.5 & 100\% & 6.9 \\
Heuristic & 92.2 & 86.6 & 72\% & 5.4 \\
RL (Uniform) & 92.4 & 87.0 & 63\% & 4.8 \\
\textbf{RL (Prioritized)} & \textbf{92.6} & \textbf{87.4} & \textbf{55\%} & \textbf{4.1} \\
\hline
\end{tabular}
}
\end{table}

Table~\ref{tab:main_results} summarizes validation performance over the voxel-based representation under different exploration strategies.

Detection rate and coverage converge across all methods under sufficient exploration, as all strategies operate on the same underlying voxel-based representation used for geometry-driven validation. Therefore, evaluation is carried out under a bounded exploration budget, where efficiency (Samples) serves as the primary differentiating factor.

The voxel-based representation is used as the reference for validation, ensuring that observed differences stem from exploration efficiency rather than variations in the validation criteria.

\begin{figure}[!ht]
\centering
\includegraphics[width=\linewidth]{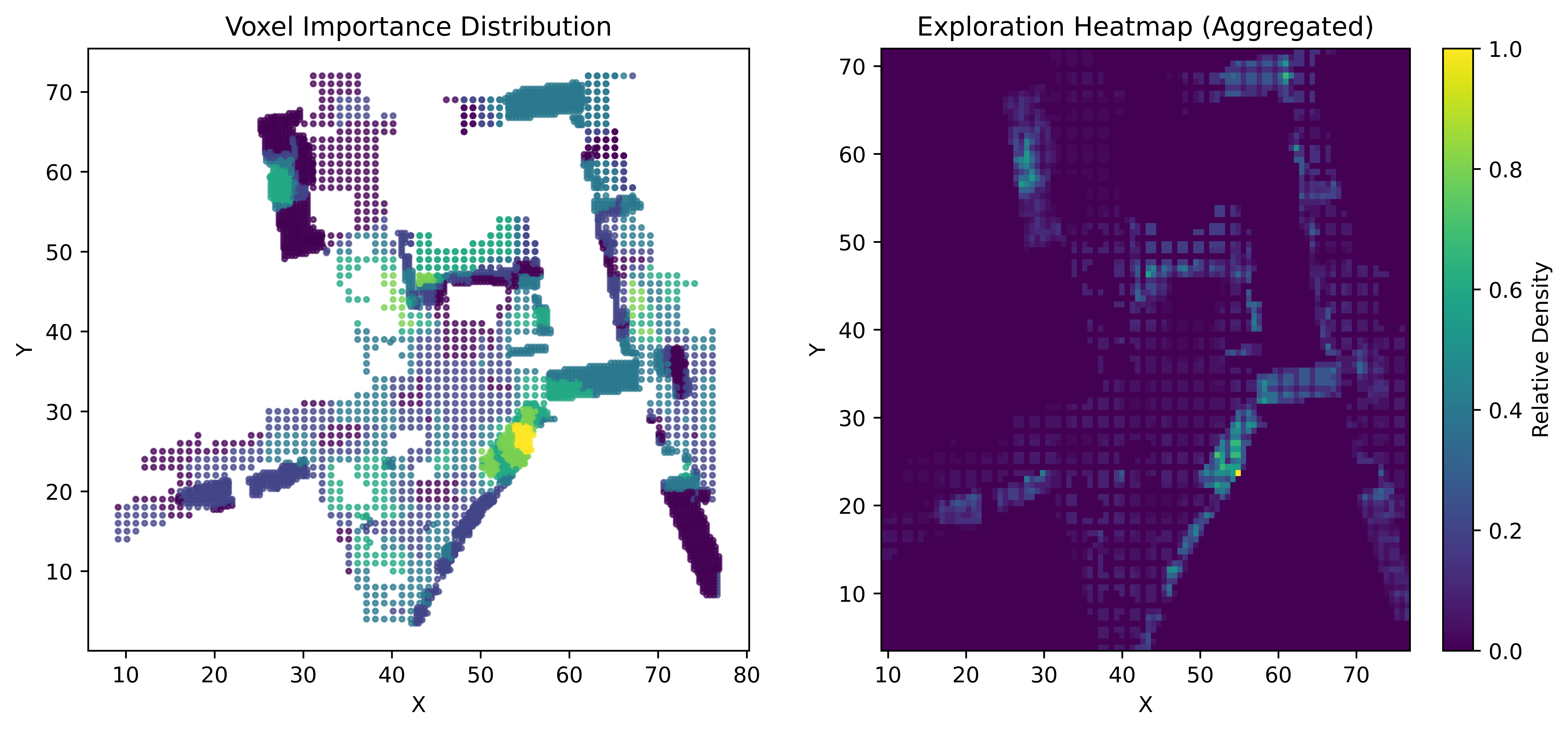}
\caption{Comparison of voxel importance distribution (left) and the corresponding aggregated heatmap (right). The heatmap highlights regions with a higher concentration of navigation-relevant features, illustrating how exploration is directed toward semantically important areas.}
\label{fig:rl_heatmap}
\end{figure}

Fig.~\ref{fig:rl_heatmap} illustrates the spatial distribution of voxel importance along with its aggregated representation. The resulting heatmap shows clear non-uniform patterns, with higher density concentrated in semantically important regions, which aligns with the observed improvements in sampling efficiency.

Across all strategies, detection rate and coverage remain consistent, while differences arise primarily in exploration efficiency. The learning-guided strategy improves sampling efficiency by prioritizing regions more likely to contain inconsistencies. The prioritized RL variant requires approximately \textbf{24\%} fewer samples than heuristic sampling and \textbf{45\%} fewer samples than exhaustive traversal (BFS/DFS) to reach similar coverage levels.

Although improvements in detection rate are modest, they are consistent across exploration strategies, reflecting the robustness of the underlying voxel-based validation. Differences in sampling efficiency directly translate to reduced validation time and computational overhead, enabling scalable deployment in automated QA workflows.

\subsection{RL Training Behavior}

The training behavior of the RL-based exploration policy is analyzed over 600 training episodes, each consisting of 2000 steps. Training is performed using a discount factor of $\gamma=0.99$, a learning rate of $1\times10^{-4}$, a replay buffer size of 30,000, and mini-batch updates.

\begin{figure}[!ht]
\centering
\includegraphics[width=0.85\linewidth]{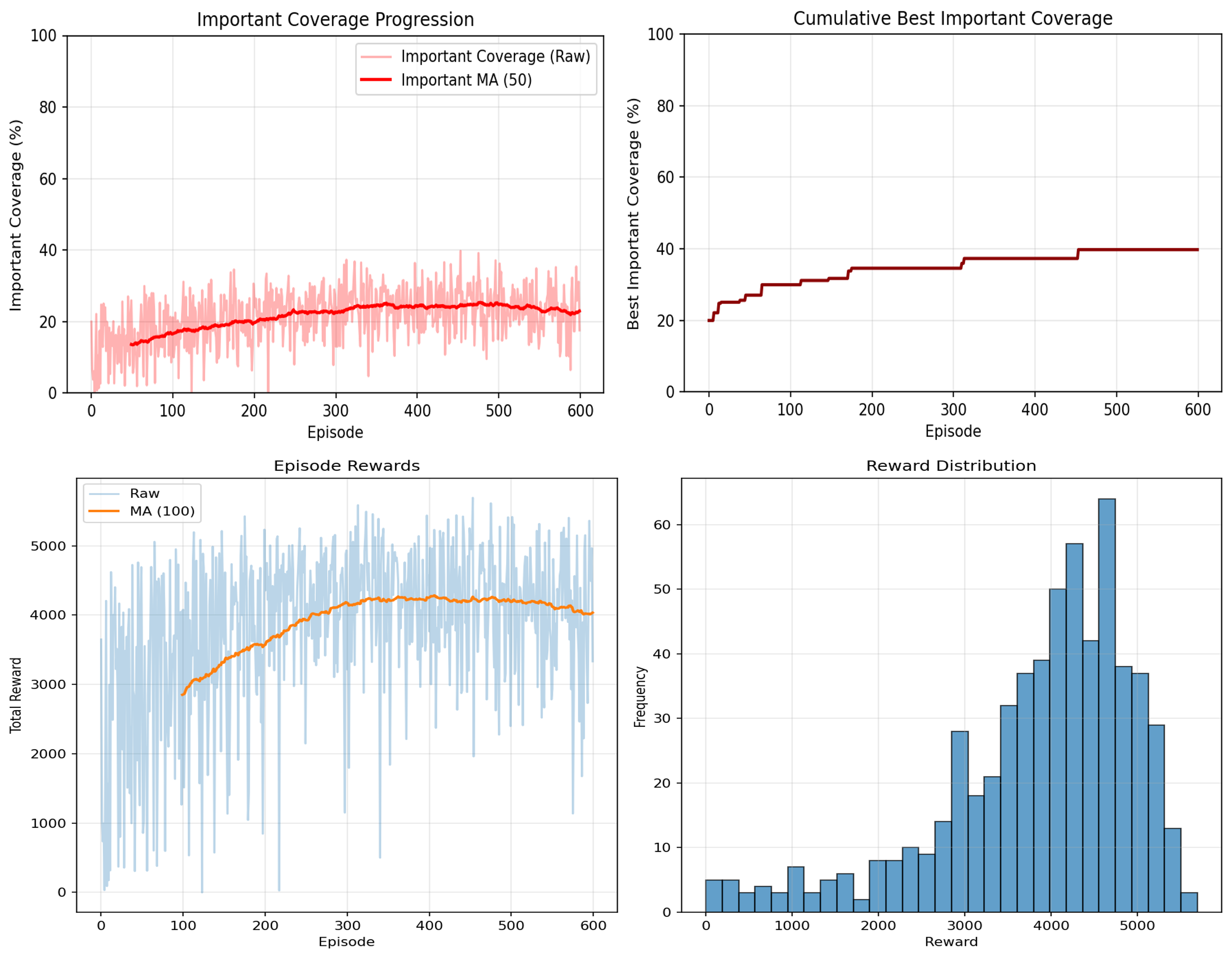}
\caption{Training progression showing convergence of the RL policy, with increasing focus on high-importance regions, followed by convergence as exploration saturates semantically important areas.}
\label{fig:rl_training_curves}
\end{figure}

Fig.~\ref{fig:rl_training_curves} shows training convergence, where the policy progressively shifts from uniform exploration to prioritizing high-importance regions. The observed $\sim$40\% coverage reflects that semantically important regions form only a subset of the voxel space, with exploration concentrated on regions most relevant for detecting navigation inconsistencies.

In later training stages, the reward signal stabilizes as the policy converges and most high-importance regions have already been explored. This plateau reflects diminishing returns from further exploration rather than a limitation of the learning process, indicating that the policy has effectively learned to prioritize defect-relevant regions.

\subsection{Ablation Study}

\begin{table}[!ht]
\centering
\small
\caption{Ablation study}
\label{tab:ablation}
\resizebox{\columnwidth}{!}{
\begin{tabular}{lccc}
\hline
Variant & Det. Rate & Cov. & Samples \\
\hline
Random (No RL) & 87.9 & 80.8 & 100\% \\
Heuristic (No RL) & 92.2 & 86.6 & 72\% \\
Low Voxel Res. & 89.8 & 83.5 & 65\% \\
\textbf{Full Method} & \textbf{92.6} & \textbf{87.4} & \textbf{55\%} \\
\hline
\end{tabular}
}
\end{table}

Removing the learning-guided exploration component reduces exploration efficiency, requiring more samples to achieve comparable detection and coverage. Heuristic sampling partially improves performance but remains less effective due to its static nature. Lower voxel resolution further reduces geometric fidelity, which in turn affects both coverage and detection accuracy. These results suggest that the main advantage of reinforcement learning comes from adaptive exploration prioritization, rather than any improvement in final detection capability.

\subsection{Discussion}

Across all evaluated environments, the framework consistently identifies navigation inconsistencies, including unreachable regions and missing walkable surfaces, by comparing geometry-derived walkability with navmesh connectivity. The voxel-based representation makes it possible to detect structural mismatches that are typically not observable through agent-based approaches.

Reinforcement learning complements this process by guiding exploration over the voxel space, reducing the number of samples needed to uncover such inconsistencies. In combination, the voxel representation and learning-guided exploration enable efficient validation without requiring exhaustive traversal.

The framework operates offline and shows consistent performance across both Unity and Anvil-based environments, with minimal integration overhead, indicating that it can be applied across a range of production scenarios.

\section{Integration into Production QA Pipelines}

The proposed framework supports automated validation workflows in large-scale game development by generating structured reports containing world-space locations of inconsistencies, affected regions, and navigation context. These reports integrate with existing debugging and visualization tools, enabling QA teams to efficiently locate, analyze, and triage navigation defects, while coverage statistics provide insight into validation completeness.

The framework operates offline and supports scalable batch processing without impacting runtime systems. Independent region processing enables parallel execution and integration into continuous integration pipelines, while semantic importance weights can be configured at inference time to prioritize validation in gameplay-critical areas such as combat zones, traversal paths, and mission-critical regions.

\section{Limitations}

The approach depends on voxel resolution and quality. While finer voxelization improves sensitivity, it increases computational cost, requiring a balance between accuracy and scalability. In smaller environments, simpler traversal strategies may perform comparably.

Semantic prioritization relies on gameplay-related metadata (e.g., patrol routes, interaction zones) and is less effective when such data is limited. The method assumes ground-based traversal and does not model multi-level navigation (e.g., climbing or vertical transitions), excluding elevated walkable regions from reconstruction.

It also assumes consistency between collision geometry and intended traversal. When collision data is simplified or misaligned with design intent, detected discrepancies may require manual interpretation.

Future work will explore hierarchical representations, multi-agent exploration, and multi-layer connectivity to improve scalability and support vertical traversal.

\section{Conclusion}

This paper presents a framework for validating navmesh consistency through geometry-driven analysis of walkable space. A voxel-based representation is used to reconstruct walkable regions from terrain and collision geometry, providing an independent reference against which the generated navmesh is compared to identify structural inconsistencies. Exploration over this voxelized space is guided using reinforcement learning, which prioritizes regions with a higher likelihood of navigation errors. Together, these components enable efficient and targeted validation, achieving high detection rates while reducing unnecessary exploration compared to uniform and heuristic strategies.

The proposed system operates offline within the game engine and integrates with existing QA workflows, supporting scalable validation across large and evolving environments. The results indicate that combining geometry-based validation with learning-guided exploration offers a practical and effective solution for automated navmesh validation in production settings.

\section*{Acknowledgments}

The authors thank the development and QA teams for their input on the navmesh validation workflows and automated testing practices, which helped shape the design and evaluation of this work. The authors also acknowledge colleagues who provided feedback on the reinforcement learning-based exploration system. 

AI-based tools were used solely for language refinement and clarity improvement and did not contribute to the research methodology, experiments, results, or technical conclusions. OpenAI’s ChatGPT was used for language refinement, while Google’s Google Gemini was used to generate simple illustrative icon elements.

\balance

\end{document}